\documentclass{jpsj2}
\title{Renormalization of Commensurate Magnetic Peak in
Ni-doped La$_{1.85}$Sr$_{0.15}$CuO$_{4}$}

\author{Masato \textsc{Matsuura}\thanks{Present address: Department of Earth and Space Science, Faculty of Science, Osaka University, Toyonaka, 560-0043.}\thanks{E-mail address: mmatsuura@ess.sci.osaka-u.ac.jp}, Maiko \textsc{Kofu}$^{1}$, Hiroyuki \textsc{Kimura}$^{1}$, and Kazuma \textsc{Hirota}$^{*}$}

\inst{Institute for Solid State Physics, University of Tokyo, Kashiwa 277-8581\\$^{1}$Institute of Multidisciplinary Research for Advanced Materials, Tohoku University, Sendai 980-8577}

\abst{We have studied the magnetic excitations 
in impurity doped La$_{1.85}$Sr$_{0.15}$Cu$_{1-y}$A$_{y}$O$_{4}$
(A=Ni or Zn) by neutron scattering.
The dispersion for Zn:$y=0.017$ is similar to that for
the impurity free sample:
incommensurate peaks with the incommensurability 
$\delta=0.12\pm0.01$ (rlu)
do not change their positions up to 21~meV.
On the other hand, for Ni:$y=0.029$,
two incommensurate peaks observed at low energies
suddenly change into 
a broad commensurate peak
at $E_\mathrm{cross}=15$~meV.
Compared to the impurity free sample
with a similar Sr-concentration $x=0.16$,
$E_\mathrm{cross}$ for Ni:$y=0.029$ is 
decreased by nearly the same factor for the reduction in $T_{c}$.
This is very similar to the shift of 
the resonance energy ($E_\mathrm{res}$) 
in Ni-doped YBa$_{2}$Cu$_{3}$O$_{7}$.
These common impurity effects on the shift of $E_\mathrm{cross}$
and $E_\mathrm{res}$ suggest the same magnetic origin
for the resonance peak in YBa$_{2}$Cu$_{3}$O$_{\delta}$
and that for a crossing point of upward and downward dispersions 
in the La$_{2-x}$Sr$_{x}$CuO$_{4}$.
We propose that the sudden change in the dispersion is 
better described by
a crossover from incommensurate spin fluctuations 
to a gapped spin wave
rather than a hourglass-like dispersion.}

\kword{high-temperature superconductor, LSCO, neutron scattering, impurity, magnetic excitation}

\begin{document}
\maketitle

\section{Introduction}

Recently, similar magnetic excitations have been 
observed in high transition temperature (high-$T_{c}$) cuprates
by inelastic neutron scattering (INS),
namely, a characteristic hourglass-like magnetic 
excitation.\cite{Arai_99,Bourges_00,Reznik_04,Stock_05,Hayden_04,Tranquada_04,Vignolle_07}
The hourglass-like excitations consist of 
downward and upward dispersions from an antiferromagnetic (AF)
zone center, the so-called ($\pi,\pi$),
and include two characteristic 
magnetic signals; low-energy incommensurate peaks
and a resonance peak.
The former signal has been observed in the two
typical high-$T_{c}$ cuprates;
YBa$_{2}$Cu$_{3}$O$_{6+\delta}$ (YBCO) and
La$_{2-x}$Sr$_{x}$CuO$_{4}$ (LSCO),
and it has been revealed that
the incommensurability $\delta$ is scaled by 
$T_{c}$ in the underdoped and optimally-doped
regions.\cite{yamada_98,dai_01}
The resonance peak is the most dominant
magnetic excitation signal in bilayer cuprates,
such as YBCO and 
Bi$_{2}$Sr$_{2}$CaCu$_{2}$O$_{8+\delta}$ 
(Bi2212).\cite{RossatMignod_91,Fong_99}
The resonance peak appears at 
${\bf Q}_{\parallel}$=($\pi,\pi$) and $\omega\sim40$~meV,
which corresponds to a crossing point of
upward and downward dispersions
in the hourglass-like dispersion.
The resonance peak in the YBCO and Bi2212 systems
grows below $T_{c}$ 
like an order-parameter,\cite{RossatMignod_91}
and the resonance energy $E_\mathrm{res}$
is scaled by $T_{c}$ ($5.8k_\mathrm{B}T_{c}$).\cite{dai_01}
Although the energy of the crossing point ($E_\mathrm{cross}$) 
in the LSCO system scales with $T_{c}$
($E_\mathrm{cross}\sim12k_\mathrm{B}T_{c}$)
in the superconducting phase,\cite{Vignolle_07,Petit_97,Hiraka_01}
the ($\pi,\pi$)-peak at the crossing point in the LSCO 
does not show any clear enhancement below 
$T_{c}$.\cite{Vignolle_07}
These experimental facts give rise to the following questions.
Are the origin of the resonance peak in YBCO and that of
the crossing point in LSCO same?
Is there any universal mechanism to determine 
$E_\mathrm{res}$ and $E_\mathrm{cross}$ 
in high-$T_{c}$ cuprates?

Impurity effects on Cu sites,
in particular, nonmagnetic Zn$^{2+}$ ($S=0$)
and magnetic Ni$^{2+}$ ($S=1$),
have been used to study a correlation 
between the magnetism and the superconductivity
because it can control $T_{c}$
without changing a carrier number
and lattice properties.
Therefore, studying impurity effects
on $E_\mathrm{cross}$
can separate magnetic and non-magnetic
contributions on the $T_{c}$-scaling
of $E_\mathrm{cross}$.
In addition,
the impurity effects of Ni and Zn
on $E_\mathrm{res}$ in YBCO
seem different:
Zn does not change $E_\mathrm{res}$
while Ni decreases $E_\mathrm{res}$
with a preserved $E_\mathrm{res}/T_{c}$ ratio.\cite{Sidis_00}
Comparing impurity effects on 
$E_\mathrm{cross}$ in LSCO and
$E_\mathrm{res}$ in YBCO
could give a clue to their 
origins.
In this report, we present 
INS measurement on impurity doped 
La$_{1.85}$Sr$_{0.15}$Cu$_{1-y}$A$_{y}$O$_{4}$
(A=Ni:$y=0.029$ and Zn:$y=0.017$).
We observed that $E_\mathrm{cross}$ for Ni:$y=0.029$
is reduced by the same factor for the reduction of $T_{c}$,
which is similar to the decrease in $E_\mathrm{res}$
for the Ni-doped YBCO.\cite{Sidis_00} 

\section{Experimental Deitals}

Single crystals were
carefully grown using a traveling-solvent 
floating zone method with a special attention
to homogeneity of the doped-impurity.
The crystals used in this report
are same as those used in ref.~15, and
the detailed sample characterizations were described
therein.
$T_{c}$, defined at midpoint of transition, drops 
from 36.8~K ($y=0$) to 11.6~K (A=Ni:$y=0.029$)
and to 16.0~K (A=Zn:$y=0.017$),
while the structural transition temperatures
from tetragonal to orthorhombic phase are nearly
same: 186 ($y=0$), 182 (Ni:$y=0.029$),
and 202~K (Zn:$y=0.017$), suggesting
the same Sr-concentration for these samples.
INS experiments were performed using the Tokyo University's
triple axis spectrometer PONTA
installed at the JRR-3 reactor in the
Japan Atomic Energy Agency.
The horizontal collimations were
40$^\prime$-40$^\prime$-80$^\prime$-80$^\prime$. 
The final neutron energy was fixed at 14.7~meV
with a pyrolytic graphite (PG) analyzer.
A PG filter was placed in front of the analyzer 
to diminish scattering for higher order neutrons.
Throughout this paper,
we label the momentum transfer $(Q_x,Q_y,Q_z)$
in units of reciprocal lattice vectors
$a^{*}\sim b^{*}=1.178$~\AA$^{-1}$ 
and $c^{*}=0.4775$~\AA$^{-1}$ for the orthogonal notaion.

\begin{figure}[htb]
\begin{center}
\includegraphics[width=8cm]{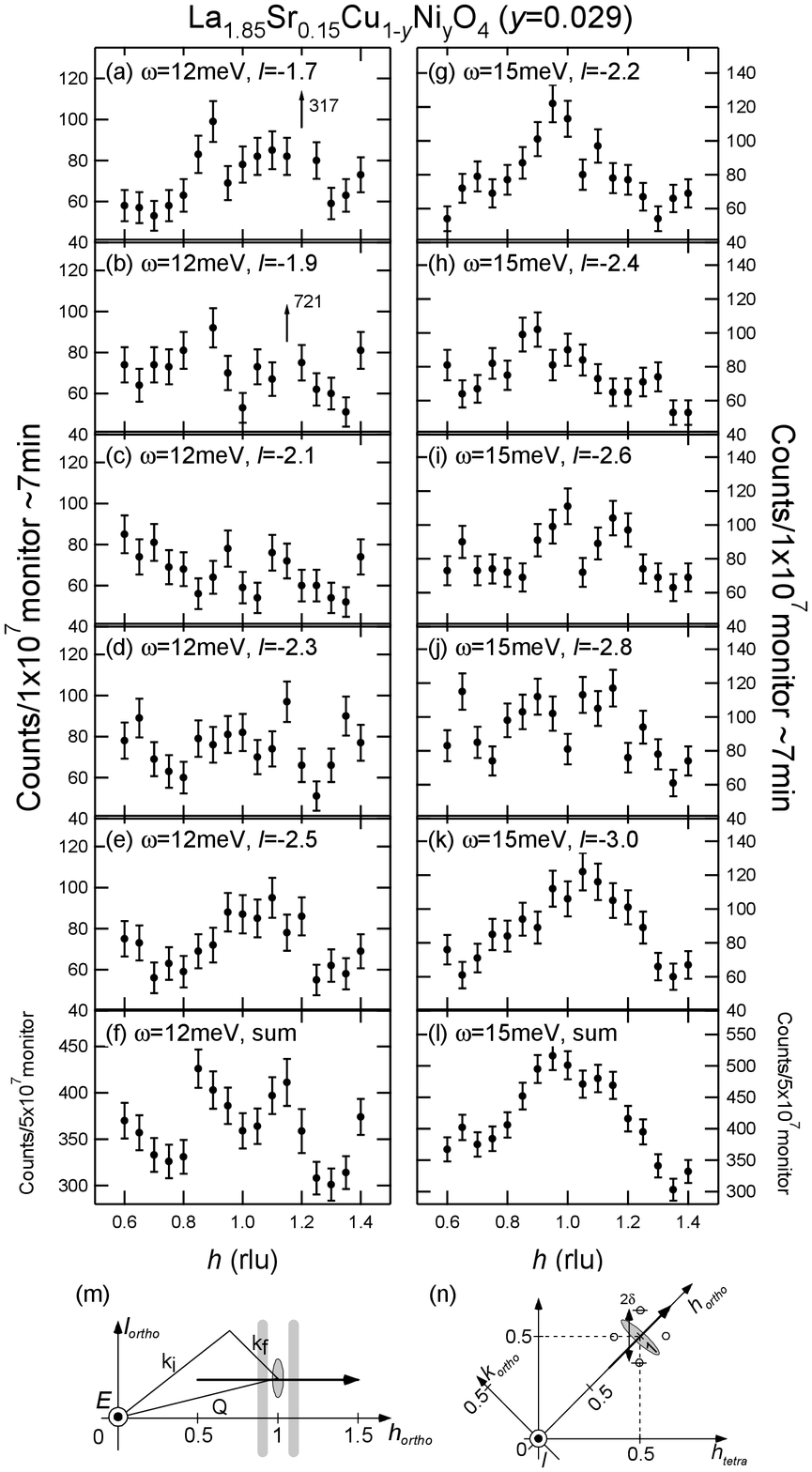}
\end{center}
\caption{Profiles of constant-$E$ scan of Ni:$y=0.029$
with energy transfer fixed at (a)-(e) 12~meV
and (g)-(k) 15~meV measured at $T=11$~K.
The trajectory of the scan is along [100]$_\mathrm{ortho}$
around the AF zone center ($10l$),
which is shown by the arrow in (m). 
To get a sharp peak, the instrumental resolution 
(gray ellipsoid) is tuned to be parallel to 
magnetic rod (two gray rods) by selecting $l$ of the scans.
At each energy, five [100]$_\mathrm{ortho}$-scans 
at different fixed $l$'s 
were measured and summed up to dilute phonon peaks and 
make magnetic signals clearer as described in the text.
The summed profiles are shown in (f) and (l) for 
$\omega=12$ and 15~meV, respectively.
(n) Wave vectors of the incommensurate peaks 
in the ($hk0$) zone in tetragonal and orthorhombic notations.
}
\label{expdetail}
\end{figure}

For the monolayer high-$T_{c}$ cuprates, 
magnetic scattering forms a rod-like signal along the $l$-direction
in the $Q$-space due to the two-dimensional spin correlation
in the real-space.
We mounted the samples in the $(h,0,l)_\mathrm{ortho}$ zone
and chose $l$ to make the instrumental resolution ellipsoid parallel
to the magnetic rod as shown in a schematic diagram
in Fig.\ref{expdetail}(m)
to get maximum magnetic signal.
Such $l$'s ($l_\parallel$'s) are 
-2.1, -2.2, -2.3, -2.6, -2.6, -2.9, -3.0, and -3.3 
for $\omega=12$, 13, 14, 15, 16, 17, 18, and 21~meV, 
respectively. 
In the three dimensional reciprocal space,
four incommensurate peaks appear at 
($\frac{1}{2}\pm\delta, \frac{1}{2},0$) and 
($\frac{1}{2},\frac{1}{2}\pm\delta,0$) 
in the tetragonal notation,
which correspond to
($1\pm\delta,\delta,0$) and 
($1\pm\delta,-\delta,0$)
in the orthorhombic notation.
Although the incommensurate peaks are not
in the $(h,0,l)_\mathrm{ortho}$ zone,
signals are detectable owing to a broad $q$-resolution
along $q_{z}$ ($k_\mathrm{ortho}$ in this case) direction
as shown in a schematic diagram
in Fig.\ref{expdetail}(n).
In addition to the tune of the instrumental resolution,
we averaged signals at several $l$'s.
Because a coherent single phonon scattering has a strong 
$l$-dependence while a magnetic signal 
has almost no $l$-dependence,
%owing to weak magnetic 
%interaction between adjacent CuO$_{2}$ planes,
this procedure 
dilutes phonon contributions and 
makes magnetic signals clearer.
All the profiles of constant-energy scans in this report 
are average of five constant-energy scans 
along the [$100$]$_\mathrm{ortho}$ 
direction around ($10l$).
Figure~\ref{expdetail} shows examples of such
$l$-average procedure
for the Ni:$y=0.029$ with energy transfer fixed at
12 and 15~meV.
The fixed five $l$'s are selected around $l_\parallel$ 
with $\pm0.2$ and $\pm0.4$ (rlu).
%as $l=-1.7$, -1.9, -2.1 ($l_\parallel$), -2.3, and -2.5 
%for $\omega=12$~meV,
%and $l=-2.2$, -2.4, -2.6 ($l_\parallel$), -2.8, and -3.0 
%for $\omega=15$~meV.
Spurious peaks such as Figs.~\ref{expdetail}(a) (1.2,0,-1.7)
and (b) (1.15,0,-1.9) were not used for this
$l$-average procedure.
Longitudinal acoustic and optical phonon 
branches along [110]$_\mathrm{tetra}$ 
in La$_{1.9}$Sr$_{0.1}$CuO$_{4}$
were reported around 14 and 18~meV
at the zone boundary, respectively.\cite{Pintschovius_91}
Although phonon or spurious peaks appear at some $l$'s,
similar peak structures,
incommensurate peaks for $\omega=12$~meV
and a single broad peak for $\omega=15$~meV,
were commonly observed 
through five $l$'s.
These peak features are clearer for the summed profiles
as shown in Figs.~\ref{expdetail}(f) and (l).

\section{Experimental Results}

\begin{figure}[htb]
\begin{center}
\includegraphics[width=7.2cm]{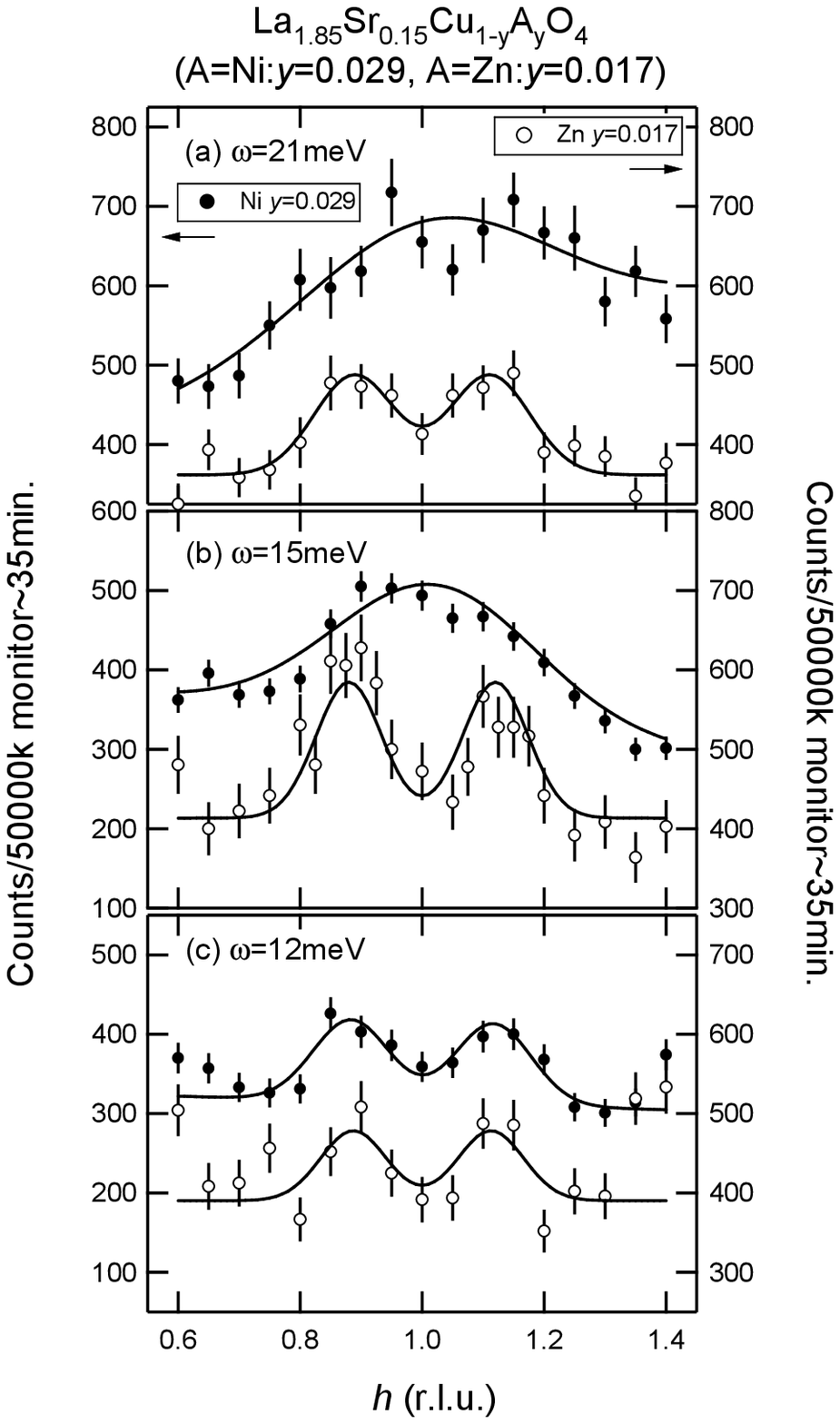}
\end{center}
\caption{(a)-(c) Profiles of constant-$E$ scan 
along the [100]$_{ortho}$ direction
around the AF zone center ($h=1.0$) at $T=11$~K
for Ni:$y=0.029$ (closed circles) 
and Zn:$y=0.017$ (solid circles).
}
\label{cnstE_Ni_Zn}
\end{figure}

Figures~\ref{cnstE_Ni_Zn} (a)-(c) show 
profile of constant-energy scans 
at $\omega=12,15,$ and 21~meV
obtained by the $l$-average procedure.
For Zn:$y=0.017$, the incommensurate peaks appear
at the same position, $h=1\pm\delta$ 
with $\delta=0.12\pm0.01$ (rlu)
for all energies, which is similar to the impurity free sample $y=0$.
Two symmetrical Gaussians reproduce the data for Zn:$y=0.017$
well even at 21~meV, where optical phonon modes are expected,
which confirms validity of the $l$-average procedure.
For Ni:$y=0.029$, at $\omega=12$~meV,
the incommensurate peaks were observed at the same $\delta=0.116\pm0.007$
as $y=0$ and Zn:$y=0.017$.
On the other hand, 
the two incommensurate peaks merge into a single broad peak
for $\omega\geq15$~meV of Ni:$y=0.029$.
Because a drastic change in lattice dynamics is hardly expected
by a small amount of impurities,
the single broad peak above $\omega\geq15$~meV
for Ni:$y=0.029$ cannot be explained by phonon contributions
which is absent for Zn:$y=0.017$.
Therefore, the changes in these profiles are
associated with a change in magnetism and/or superconductivity
by Ni-doping.

\begin{figure}[htb]
\begin{center}
\includegraphics[width=8.5cm]{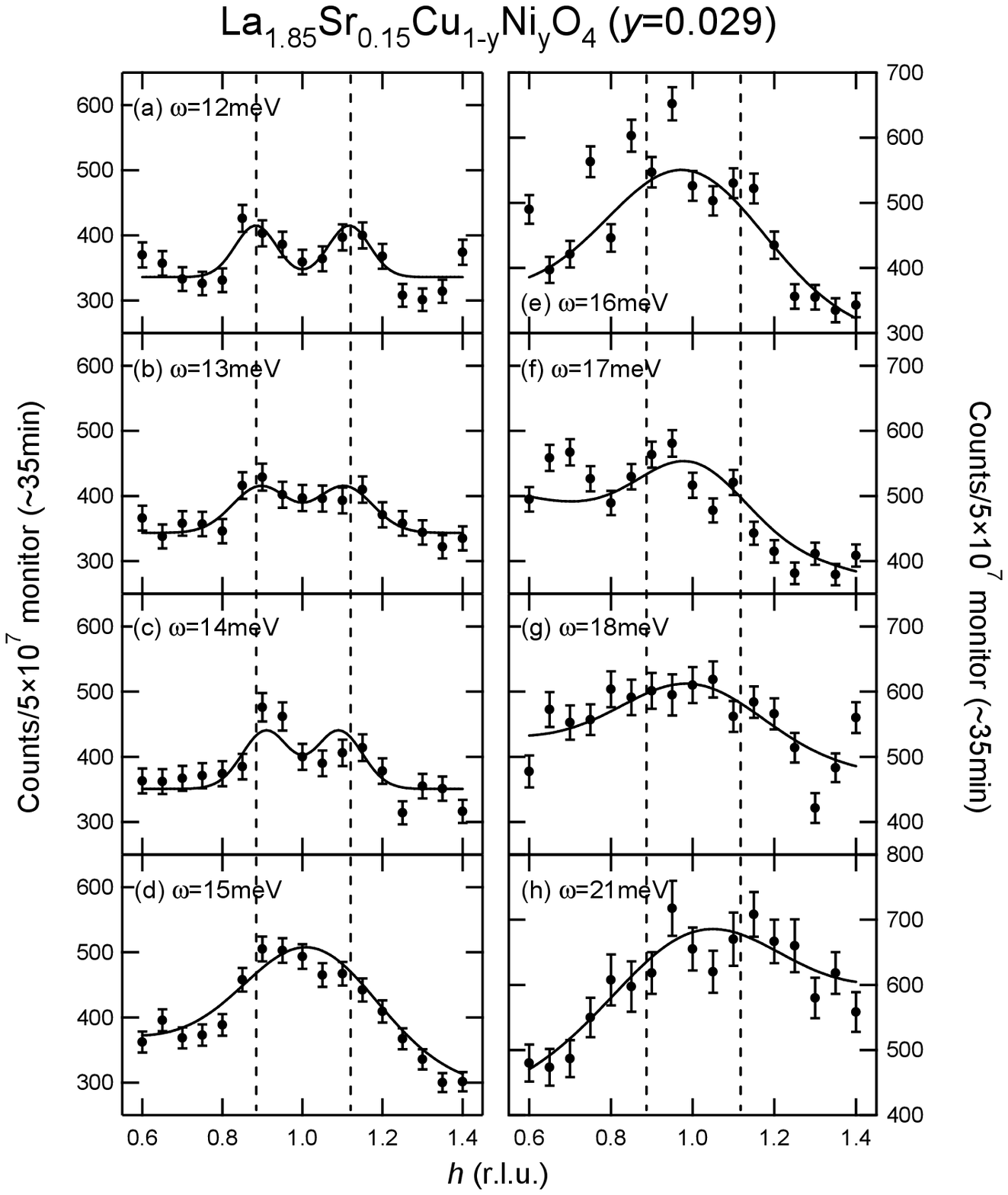}
\end{center}
\caption{(a)-(c) Profiles of $l$-averaged 
constant-energy scans at several energies
for Ni:$y=0.029$ measured at $T=11$~K.
The trajectory of the scan is same as
Fig.~\ref{expdetail}.
Solid lines show fit to two symmetrical Gaussian functions
around $h=1.0$ for $\omega\leq14$~meV and to a single Gaussian 
one at $h=1.0$ for $\omega\geq15$~meV on a sloped background.
The vertical broken lines show the positions
of the incommensurate peaks at $\omega=12$~meV.
}
\label{cnstE_Ni}
\end{figure}

To explore the in-plane dispersion for Ni:$y=0.029$,
constant-energy scans were carried out at several energies
at $T=11$~K as shown in Figs.~\ref{cnstE_Ni}.
%Typical energy resolution of these scans were about 1~meV.
For $12\leq \omega\leq14$~meV,
the incommensurability $\delta$ varies little
with increasing energy.
At $\omega=15$~meV,
the scattering intensity around the AF zone center
increases, and the two incommensurate peaks suddenly
become a single broad peak.
Above 16~meV,
although background becomes high
owing to optical phonons at 18 and 21~meV,\cite{Boni_88}
a broad peak was commonly observed at the
AF zone center.

\begin{figure}[tb]
\begin{center}
\includegraphics[width=8.5cm]{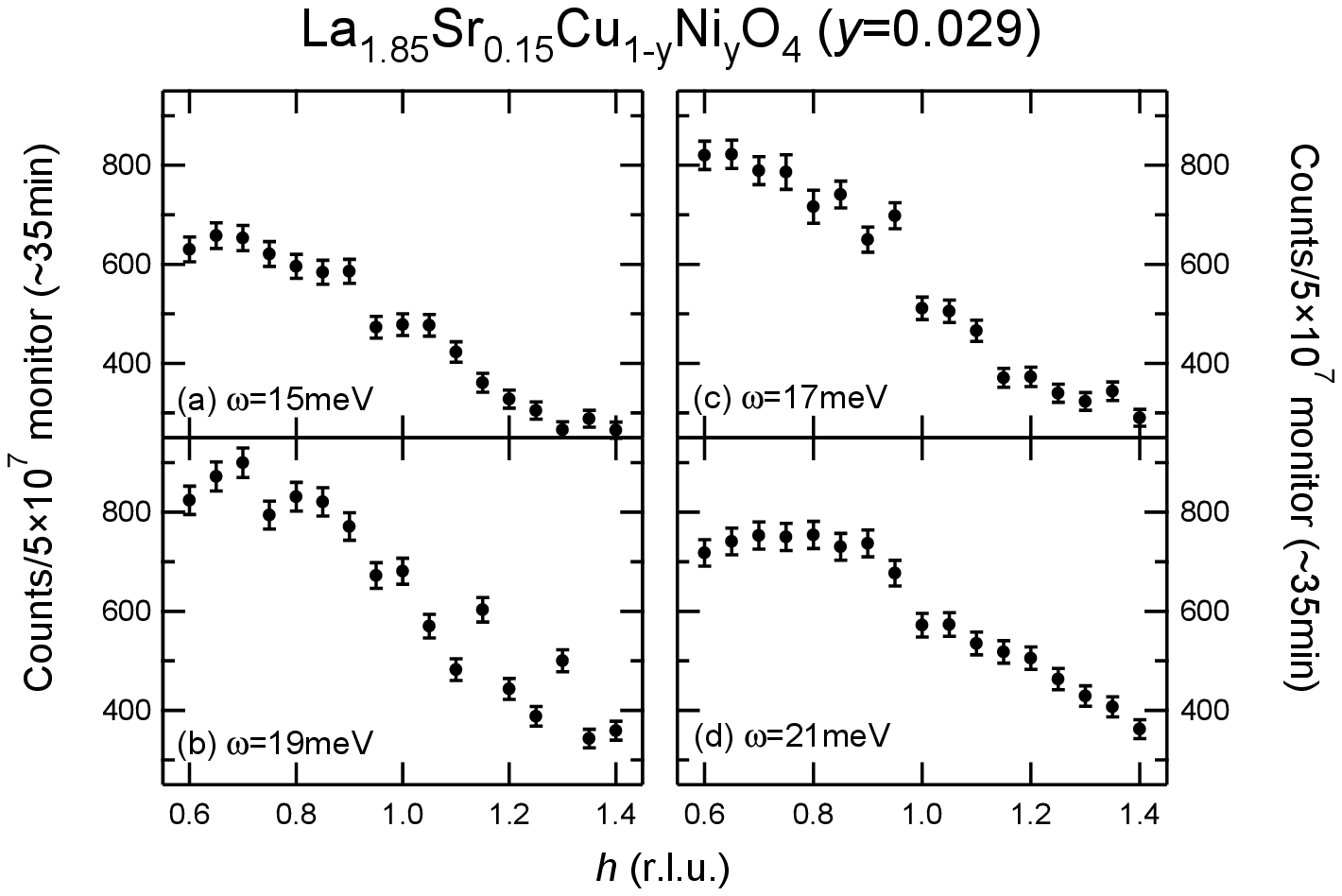}
\end{center}
\caption{Profiles of $l$-averaged 
constant-energy scans for Ni:$y=0.029$ 
measured at $T=300$~K.
The scans are same as those
in Fig.~\ref{cnstE_Ni}.
}
\label{cnstE_Ni_RT}
\end{figure}

Figure~\ref{cnstE_Ni_RT} shows 
the same constant-energy scans
measured at $T=300$~K.
All profiles show monotonical
decrease with increasing $h$ and 
have no appreciable peak structure.
Therefore, the broad peak structure
observed at $T=11$~K above $\omega\geq 15$~meV
disappears at high temperatures,
indicating magnetic origin of these peaks.

\section{Discussion}

\begin{figure}[htb]
\begin{center}
\includegraphics[width=8.5cm]{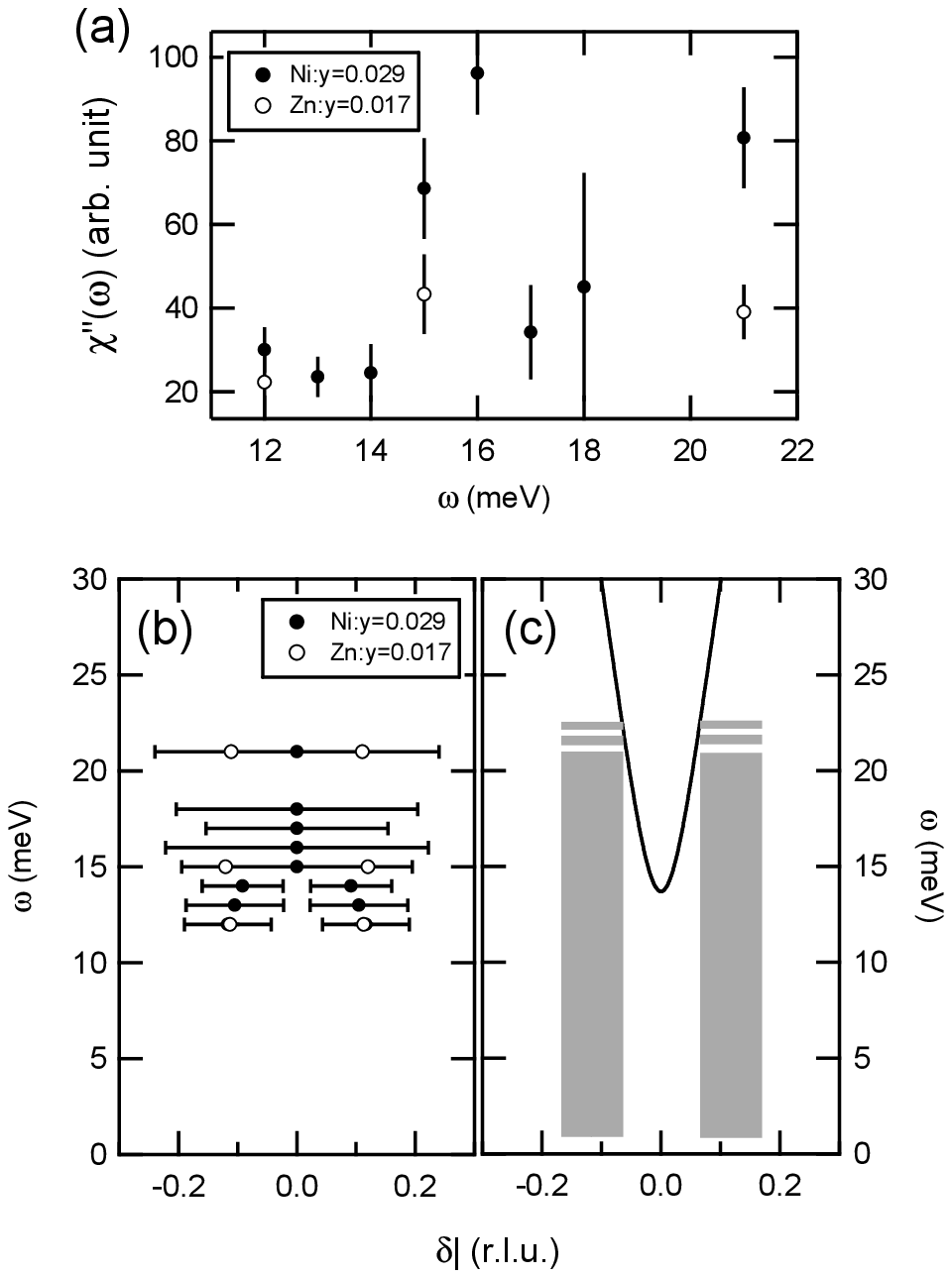}
\end{center}
\caption{(a) The $Q$-integrated susceptibility 
$\chi^{\prime\prime}$($\omega$) and (b) dispersion
measured at $T=11$~K.
The solid and open circles represent
the data for Ni:$y=0.029$ and 
Zn:$y=0.017$, respectively.
(c) Schematic diagram of the dispersionless
incommensurate fluctuations 
and spin wave dispersion (described in text).}
\label{disp}
\end{figure}

The $Q$-integrated susceptibility 
$\chi^{\prime\prime}$($\omega$)
for Ni:$y=0.029$ and Zn:$y=0.017$
are shown in Fig.~\ref{disp}(a).
$\chi^{\prime\prime}$($\omega$) for Ni:$y=0.029$
exhibits a maximum near 15-16~meV
at which the two incommensurate
peaks merge into the broad commensurate 
peak at ($\pi$,$\pi$).
The maximum $\chi^{\prime\prime}$($\omega$)
at $E_\mathrm{res}$ or $E_\mathrm{cross}$ is
commonly observed in high-$T_{c}$
cuprates.
Together with a change
from the incommensurate peak structure
to the commensurate one,
we conclude that $E_\mathrm{cross}$ for Ni:$y=0.029$
is 15-16~meV, which is 2.7 times smaller than
$E_\mathrm{cross}$ for the 
impurity free LSCO with similar Sr-concentration
$x=0.16$.\cite{Vignolle_07}
Hiraka {\it et al.} revealed that doped Ni-ions 
form a strong bound state with holes
and reduce the effective hole carrier number,
e.g.
$x=0.15$ with $y=0.029$ corresponds to
$x=0.121$ with $y=0$.~\cite{Hiraka_07}
However, the reduction in $E_\mathrm{cross}$ can not be 
explained only by reduction in the hole carrier,
because $E_\mathrm{cross}$ of 25~meV for $x=0.10$\cite{Petit_97}
is even higher than the current result on Ni:$y=0.029$.
Instead,
$E_\mathrm{cross}$ decreases with a preserved
$E_\mathrm{cross}/T_{c}$ ratio.
% in other words, $E_\mathrm{cross}$ is scaled by $T_{c}$.
%$E_\mathrm{cross}$($y=0.029$)/$E_\mathrm{cross}$($y=0.0$)=0.37
%is similar to $T_{c}$($y=0.029$)/$T_{c}$($y=0.0$)=0.32.
%(36.8~K for $y=0$ and 11.6~K for Ni:$y=0.029$).

$\chi^{\prime\prime}$($\omega$) and dispersion
for Zn:$y=0.017$ are shown in Figs.~\ref{disp}(a) and (b)
as open circles.
The incommensurability $\delta$
is same as impurity free sample
and is constant up to 21~meV,
which indicates that Zn-doping does not affect
$E_\mathrm{cross}$.
Sidis {\it et al.} have studied the
impurity (Ni and Zn) effect on $E_{res}$ 
in optimally doped YBCO, and found that
the resonance peak 
shifts to lower energy by Ni-doping
with a preserved $E_\mathrm{res}/T_{c}$ ratio
while Zn-doping does not change $E_\mathrm{res}$.\cite{Sidis_00}
These similar impurity effects on 
$E_\mathrm{cross}$ and $E_\mathrm{res}$
indicate a common origin for the 
crossing point peak in LSCO 
and the resonance peak in YBCO.

The marked difference between the LSCO 
and the YBCO families is the thermal
variation of the magnetic spectrum
at $E_\mathrm{cross}$ or $E_\mathrm{res}$;
an order-parameter like behavior
only for the resonance peak.
However, for the Zn-doped optimum YBCO,
the intensity of the resonance peak
gradually decreases with increasing temperature,
and a substantial spectral weight remains
above $T_{c}$.\cite{Sidis_00}
Recently, Koike {\it el al.} have 
revealed a small superconducting volume fraction
in a wide doping range of LSCO
due to phase separation
and/or spatial inhomogeneity.\cite{Koike_08}
Since Zn is known to induce localized moments
and reduce superconducting region,
the LSCO system seems to be in similar 
situation as Zn-doped YBCO.
Therefore, 
we speculate that the different temperature dependence
of the magnetic peak between LSCO and YBCO
is associated with
the spacial inhomogeneity of CuO$_{2}$ plane
in the LSCO system.

If the resonance peak and the crossing point peak
come from the same origin,
the next question is ``what determines
$E_\mathrm{cross}$ and $E_\mathrm{res}$?''
Several theoretical\cite{Batista_01,Kruger_03} and
experimental\cite{Kofu_condmat,Matsuda_08} studies 
claimed that $E_\mathrm{cross}$
in the LSCO system depends on $\delta$ 
assuming the same downward-dispersion.
In this view, the crossing point corresponds
to the top of the downward dispersion.
This model explains large $E_\mathrm{cross}\sim55$~meV
of the stripe ordered La$_{1.875}$Ba$_{0.125}$CuO$_{4}$
by a large $\delta=0.14$.
However, the small $E_\mathrm{cross}$ (15~meV) for Ni:$y=0.029$
in spite of almost the same $\delta$ ($\delta\sim0.12$)
as that of the impurity free sample 
indicates $\delta$ is not a good parameter to
control $E_\mathrm{cross}$.

On the other hand,
the crossing point could be viewed as
the bottom of the the upward dispersion.
Tranquada {\it et al.} analyzed upward branch
observed in stripe ordered LBCO
as a spin wave from a two-leg spin ladder.\cite{Tranquada_04}
Barnes {\it et al.} have calculated the dispersion 
of two-leg spin ladder by a $t-J$ model, and have shown that
%\begin{equation}
%E(q)=[E_{\mathrm{zb}}^{2}\sin^{2}(\pi q)+E_{\mathrm{gap}}^{2}\cos^{2}(\pi q)+c^{2}\sin^{2}(2\pi q)]^{\frac{1}{2}}
%\end{equation}
%Here, $q$ is the reduced wave vector from ($\pi,\pi$), 
%$E_\mathrm{zb}$ and $E_\mathrm{gap}$ are the energies of 
%spin wave excitations at zone boundary and gap energy,
%and c is constant.
a spingap ($E_\mathrm{gap}$) is proportional to $J$
and $J_{\perp}/J$.\cite{Barnes_93}
Here, $J$ is the magnetic interaction
along spin-chain $J$,
and $J_{\perp}$ is that between spin-chains.
$E_\mathrm{gap}$ monotonically increases from 0 
with increasing $J_{\perp}/J$,
and reaches $J/2$ at $J=J_{\perp}$.
In this model, $E_\mathrm{cross}$ can be controlled
by changing $J$ and $J_{\perp}/J$.
A Cu NQR study on Ni-doped optimum YBCO
has shown that $1/T_{1}$, 
that is associated with $\chi^{\prime\prime}$,
is scaled to $T_{c}$, indicating the energy scale 
of magnetic excitation, $J$, is scaled by $T_{c}$
for Ni-doping.~\cite{Tokunaga_97}
Therefore, we speculate that the change in $J$
by Ni-doping affects $E_\mathrm{cross}$ 
and $E_\mathrm{res}$ as well as $T_{c}$.
Upon Sr-doping in La$_{2-x}$Sr$_{x}$CuO$_{4}$, 
$J$ is known to decrease from 146~meV ($x=0$)\cite{Coldea_01}
to 81~meV ($x=0.16$)\cite{Vignolle_07},
while $E_\mathrm{cross}$ increases linearly with $x$ 
and reaches $J/2\sim41$~meV at $x=0.16$.\cite{Matsuda_08}
This might be explained by $J_{\perp}<J$ in the underdoped
region and $J_{\perp}=J$ at the optimum doping.

The solid line in Fig.~\ref{disp}(c) shows a
calculated dispersion of the spin-ladder model
with $J=27$~meV.
Here, we simulate a spin wave dispersion
with one third of $J$ in the impurity 
free sample with a similar Sr-concentration\cite{Vignolle_07}
assuming both $E_\mathrm{cross}$ and $J$ 
are reduced with the same ratio by Ni-doping.
The full-width-at-half maximum (FWHM) 
of the single peak for $15\leq\omega\leq 21$~meV,
shown by the horizontal bars in Fig.~\ref{disp}(b),
are much broader than the calculated spin wave 
dispersion. 
Because incommensurate peaks as sharp as 
those of the impurity free sample
have been observed at $\omega=3$~meV in
Ni:$y=0.029$,\cite{Kofu_05}
the peak-broadening at $\omega=15$~meV
can not be explained only by impurities.
%whereas they are comparable to 
%2$\delta$ at lower energies.
The $Q$-profiles suddenly change from incommensurate peaks
to a commensurate peak at $\omega=15$~meV.
Since energy resolutions of these scans are about 1~meV, 
it is unlikely that the 
incommensurate peaks steeply disperse inward
at $\omega=15$~meV.
Instead,
the profile above 15~meV is more naturally explained
by the sum of incommensurate peaks with the same $\delta$
as that at low energies and the commensurate
peak at ($\pi,\pi$).
We thus propose a new view of the magnetic excitations
in high-$T_{c}$ cuprates: a crossover from
incommensurate peaks
to a gapped spin wave excitation
shown as a schematic diagram in
Fig.~\ref{disp}(c).

Vignolle {\it et al.} have revealed a two-component
structure of magnetic excitations in optimum LSCO
from two peaks in $\chi^{\prime\prime}$($\omega$).\cite{Vignolle_07}
The low energy component, that comes from 
the incommensurate peaks, has a peak around 18~meV
and rapidly decrease above 20~meV,
whereas the high energy component exhibits
a peak near 40~meV.
If the energy scale of low energy component 
is renormalized by Ni-doping as high energy component,
incommensurate peaks should be attenuated
above 6-7~meV.
The constant intensity of the incommensurate peaks
at least up to 14~meV means that
Ni-doping differently renormalizes the energy scales
of two components.
This suggests different origins for the 
incommensurate peaks and a gapped spin wave excitation.

Similar incommensurate to a commensurate response
has been observed in the paramagnetic state of 
Cr.\cite{Fukuda_96}
In Cr, 
the incommensurate peak shifts inward
when a crossover occurs from the coexisting state 
of the commensurate and incommensurate components 
to commensurate one at 30-40~meV.
We speculate that
the downward dispersion below 
$E_\mathrm{cross}$ and $E_\mathrm{res}$
observed in the impurity free samples
is explained by crossover of two components;
incommensurate spin fluctuations and a gapped 
spin-wave from ($\pi,\pi$).

In conclusion,
we observed that magnetic excitations change 
from incommensurate to a commensurate response
at $E_\mathrm{cross}=15$~meV for the magnetic 
impurity-doped La$_{1.85}$Sr$_{0.15}$Cu$_{1-y}$O$_{4}$
(Ni:$y=0.029$).
Similar impurity effect on the reduction 
in $E_\mathrm{cross}$ and $E_\mathrm{res}$
with a preserved ratio to $T_{c}$ suggests
the same origin for the crossing point of
hourglass-like dispersion in LSCO
and the resonance peak in YBCO.
On the other hand, no spectral shift to
low energy was observed for the incommensurate peaks,
indicating different origins for the two components;
incommensurate spin fluctuations at low energies
and a dispersive mode at high energy region.
We propose a new picture of magnetic excitation
in high-$T_{c}$ superconductor;
a crossover from incommensurate spin fluctuations 
to a gapped spin wave
rather than the hourglass-like dispersion.

\section*{Acknowledgments}
The INS experiments at PONTA was performed under 
the joint-research program of ISSP, the University of Tokyo.
MM was supported by a Grant-In-Aid for 
Encouragement of Young Scientists (B) (17740217, 2004) from the Japanese
Ministry of Education, Science, Sports and Culture.


\begin{thebibliography}{99} %% The number "99" means that this list has more than nine items.
\bibitem{Arai_99} M. Arai, T. Nishijima, Y. Endoh, T. Egami, S. Tajima, K. Tomimoto, Y. Shiohara, M. Takahashi, A. Garrett, and S. M. Bennington: Phys. Rev. Lett. {\bf 83} (1999) 608.
\bibitem{Bourges_00} P. Bourges, Y. Sidis, H. F. Fong, L. P. Regnault, J. Bossy, A. Ivanov, and B. Keimer: Science {\bf 288} (2000) 1234.
\bibitem{Reznik_04} D. Reznik, P. Bourges, L. Pintschovius, Y. Endoh, Y. Sidis, T. Masui, and S. Tajima: Phys. Rev. Lett. {\bf 93} (2004) 207003.
\bibitem{Stock_05} C. Stock, W. J. L. Buyers, R. A. Cowley, P. S. Clegg, R. Coldea, C. D. Frost, R. Liang, D. Peets, D. Bonn, and W. N. Hardy: Phys. Rev. B {\bf 71} (2005) 24522.
\bibitem{Hayden_04} S. M. Hayden, H. A. Mook, Pengcheng Dai, T. G. Perring and F. Dogan: Nature {\bf 429} (2004) 531.
\bibitem{Tranquada_04} J. M. Tranquada, H. Woo, T. G. Perring, H. Goka, G. D. Gu, G. Xu, M. Fujita, and K. Yamada: Nature {\bf 429} (2004) 534.
\bibitem{Vignolle_07} B. Vignolle, S. M. Hayden, D. F. McMorrow, H. M. R$\o$nnow, B. Lake, C. D. Frost, and T. G. Perring: Nature Physics {\bf 3} (2007) 163.
\bibitem{yamada_98} K. Yamada, C. H. Lee, K. Kurahashi, J. Wada, S. Wakimoto, S. Ueki, Y. Kimura, Y. Endoh, S. Hosoya, and G. Shirane: Phys. Rev. B {\bf 57} (1998) 6165.
\bibitem{dai_01} P. Dai, H. A. Mook, R. D. Hunt, and F. Do$\rm\breve{g}$an: Phys. Rev. B {\bf 63} (2001) 54525.
\bibitem{RossatMignod_91} J. Rossat-Mignod, L. P. Regnault, C. Vettier, P. Bourges, P. Burlet, J. Bossy, J. Y. Henry, and G. Lapertot: Physica C {\bf 185-189} (1991) 86.
\bibitem{Fong_99} H. F. Fong, B. Bourges, Y. Sidis, L. P. Regnault, A. Ivanov, G. D. Gu, N. Koshizuka, and B. Keimer: Nature \textbf{398} (1999) 588.
\bibitem{Petit_97} S. Petit, A. H. Moudden, B. Hennion, A. Vietkin, and A. Revcolevschi: Physica B {\bf 234-236} (1997) 800.
\bibitem{Hiraka_01} H. Hiraka, Y. Endoh, M. Fujita, Y. S. Lee, J. Kulda, A. Ivanov, and R. J. Birgeneau: J. Phys. Soc. Jpn {\bf 70} (2001) 853.
\bibitem{Sidis_00} Y. Sidis, P. Bourges, H. F. Fong, B. Keimer, L. P. Regnault, J. Bossy, A. Ivanov, B. Hennion, P. Gautier-Picard, G. Collin, D. L. Millius, and I. Aksay: Phys. Rev. Lett. {\bf 84} (2000) 5900.
\bibitem{Kofu_05} M. Kofu, H. Kimura, and K. Hirota: Phys. Rev. B {\bf 72} (2005) 064502.
\bibitem{Pintschovius_91}
L. Pintschovius, N. Pyka,  W. Reichardt, A. Yu. Rumiantsev, N. L. Mitrofanov, A. S. Ivanov, G. Collin, P. Bourges: Physica B {\bf 174} (1991) 323.
\bibitem{Boni_88} P. B\"oni, J. D. Axe, G. Shirane, R. J. Birgeneau, D. R. Gabbe, H. P. Jenssen, M. A. Kastner, C. J. Peters, P. J. Picone, and T. R. Thurston: Phys. Rev. B {\bf 38} (1988) 185.
\bibitem{Hiraka_07} H. Hiraka, S. Ohta, S. Wakimoto, M. Matsuda, and K. Yamada: J. Phys. Soc. Jpn. {\bf 76} (2007) 074703.
\bibitem{Koike_08} Y. Koike, T. Adachi, Y. Tanabe, K. Omori, T. Noji, and H. Sato: J. of Physics {\bf 108} (2008) 012003.
\bibitem{Batista_01} C. D. Batista, G. Ortiz, and A. V. Balatsky: Phys. Rev. B {\bf 64} (2001) 172508.
\bibitem{Kruger_03} F. Kr\"uger, and S. Scheidl:  Phys. Rev. B {\bf 67} (2003) 134512.
\bibitem{Kofu_condmat} M. Kofu, T. Yokoo, F. Trouw, and K. Yamada: cond-mat/0710.5766.
\bibitem{Matsuda_08} M. Matsuda, M. Fujita, S. Wakimoto, J. A. Fernandez-Baca, J. M. Tranquada,  K. Yamada: Phys. Rev. Lett. {\bf 101} (2008) 197001.
\bibitem{Barnes_93} T. Barnes, E. Dagotto, J. Riera, E. S. Swanson: Phys. Rev. B {\bf 47} (1993) 3196.
\bibitem{Tokunaga_97} Y. Tokunaga, K. Ishida, Y. Kitaoka, and K. Asayama: Solid State Commun. {\bf 103} (1997) 43.
\bibitem{Coldea_01} R. Coldea, S. M. Hayden, G. Aeppli, T. G. Perring, C. D. Frost, T. E. Mason, S. W. Cheong, and Z. Fisk: Phys. Rev. Lett. {\bf 86} (2001) 5377.
\bibitem{Fukuda_96} T. Fukuda, Y. Endoh, K. Yamada, M. Takeda, S. Itoh, M. Arai, and T. Otomo: J. Phys. Soc. Jpn. {\bf 65} (1996) 1418.
\end{thebibliography}
\end{document}